\documentclass[aps,prd,draft]{revtex4}
\usepackage[latin1]{inputenc}
\usepackage{amsmath}
\usepackage{amsfonts}
\usepackage{amssymb}

\input{tcilatex}

\begin{document}

\title{Varying Couplings in Electroweak Theory}
\author{Douglas J. Shaw}
\affiliation{DAMTP, Centre for Mathematical Sciences, University of Cambridge,
Wilberforce Road, Cambridge CB3 OWA, UK}
\author{John D. Barrow}
\affiliation{DAMTP, Centre for Mathematical Sciences, University of Cambridge,
Wilberforce Road, Cambridge CB3 OWA, UK}
\date{\today}

\begin{abstract}
We extend the theory of Kimberly and Magueijo for the spacetime variation of
the electroweak couplings in the unified Glashow-Salam-Weinberg model of the
electroweak interaction to include quantum corrections. We derive the
effective quantum-corrected dilaton evolution equations in the presence of a
background cosmological matter density that is composed of weakly
interacting and non-weakly-interacting non-relativistic dark-matter
components.
\end{abstract}

\maketitle

\section{\protect\bigskip Introduction}

The studies of relativistic fine structure in the absorption lines of dust
clouds around quasars by Webb et al., \cite{Webb:2001, Murphy:2001,Webb:1999}%
, have led to widespread theoretical interest in the question of whether the
fine structure constant, $\alpha _{em}=e^{2}/\hbar c$, has varied in time
and, if so, how to accommodate such a variation by a minimal perturbation of
existing theories of electromagnetism.\ These astronomical studies have
proved to be more sensitive than laboratory probes of the constancy of the
fine structure 'constant', which currently give bounds on the time variation
of $\dot{\alpha}_{em}/\alpha _{em}\equiv -0.4\pm 16\times 10^{-16}yr^{-1}$, 
\cite{lab1}, $\left\vert \dot{\alpha}_{em}/\alpha _{em}\right\vert \
<1.2\times 10^{-15}yr^{-1}$, \cite{lab2}, $\dot{\alpha}_{em}/\alpha
_{em}\equiv -0.9\pm 2.9\times 10^{-16}yr^{-1}$, \cite{lab3} by comparing
atomic clock standards based on different sensitive hyperfine transition
frequencies, and $\dot{\alpha}_{em}/\alpha _{em}\equiv -0.3\pm 2.0\times
10^{-15}yr^{-1}$ from comparing two standards derived from 1S-2S transitions
in atomic hydrogen after an interval of 2.8 years \cite{lab4}. The quasar
data analysed in refs. \cite{Webb:2001, Murphy:2001,Webb:1999} consists of
three separate samples of Keck-Hires observations which combine to give a
data set of 128 objects at redshifts $0.5<z<3$. The many-multiplet technique
finds that their absorption spectra are consistent with a shift in the value
of the fine structure constant between these redshifts and the present of $%
\Delta \alpha _{em}/\alpha _{em}\equiv \lbrack \alpha _{em}(z)-\alpha
_{em}]/\alpha _{em}=-0.57\pm 0.10\times 10^{-5},$ where $\alpha _{em}\equiv $
$\alpha _{em}(0)$ is the present value of the fine structure constant \cite%
{Webb:2001, Murphy:2001,Webb:1999}. Extensive analysis has yet to find a
selection effect that can explain the sense and magnitude of the
relativistic line-shifts underpinning these deductions. Further
observational studies have been published in refs. \cite{chand1,chand2}
using a different but smaller data set of 23 absorption systems in front of
23 VLT-UVES quasars at $0.4\leq z\leq 2.3$ and have been analysed using an
approximate form of the many-multiplet analysis techniques introduced in
refs. \cite{Webb:2001, Murphy:2001,Webb:1999}. They obtained $\Delta \alpha
_{em}/\alpha _{em}\equiv -0.6\pm 0.6\times 10^{-6}$; a figure that disagrees
with the results of refs. \cite{Webb:2001, Murphy:2001,Webb:1999}. However,
reanalysis is needed in order to understand the accuracy being claimed.
Other observational studies of lower sensitivity have also been made using
OIII emission lines of galaxies and quasars. The analysis of data sets of 42
and 165 quasars from the SDSS gave the constraints $\Delta \alpha
_{em}/\alpha _{em}\equiv 0.51\pm 1.26\times 10^{-4}$ and $\Delta \alpha
_{em}/\alpha _{em}\equiv 1.2\pm 0.7\times 10^{-4}$ respectively for objects
in the redshift range $0.16\leq z\leq 0.8$ \cite{sdss}. Observations of a
single quasar absorption system at $z=1.15$ by Quast et al \cite{qu} gave $%
\Delta \alpha _{em}/\alpha _{em}\equiv -0.1\pm 1.7\times 10^{-6}$ , and
observations of an absorption system at $z=1.839$ by Levashov et al \cite%
{lev} gave $\Delta \alpha _{em}/\alpha _{em}\equiv 4.3\pm 7.8\times 10^{-6}$%
. A preliminary analysis of constraints derived from the study of the OH
microwave transition from a quasar at $z=0.2467$, a method proposed by
Darling \cite{darl}, has given $\Delta \alpha _{em}/\alpha _{em}\equiv
0.51\pm 1.26\times 10^{-4}$, \cite{oh}$.$A comparison of redshifts measured
using molecules and atomic hydrogen in two cloud systems by Drinkwater et al 
\cite{drink} at $z=0.25$ and $z=0.68$ gave a bound of $\Delta \alpha
_{em}/\alpha _{em}<5\times 10^{-6}$ and an upper bound on spatial variations
of $\delta \alpha _{em}/\alpha _{em}<3\times 10^{-6}$ over 3 Gpc at these
redshifts. A new study comparing UV absorption redshifted into the optical
with redshifted 21cm absorption lines from the same cloud in a sample of 8
quasars by Tzanavaris et al \cite{tz}. This comparison probes the constancy
of $\alpha ^{2}g_{p}m_{e}/m_{p}$ and gives $\Delta \alpha _{em}/\alpha
_{em}\equiv 0.18\pm 0.55\times 10^{-5}$ if we assume that the
electron-proton mass ratio and proton $g$-factor, $g_{p}$, are both constant.

Observational bounds derived from the microwave background radiation
structure \cite{cmb} and Big Bang nucleosynthesis \cite{bbn} are not
competitive at present (giving $\Delta \alpha _{em}/\alpha _{em}\lesssim
10^{-2}$ at best at $z\sim 10^{3}$ and $z\sim 10^{9}-10^{10}$) with those
derived from quasar studies, although they probe much higher redshifts.

Other bounds on the possible variation of the fine structure constant have
been derived from geochemical studies, although they are subject to awkward
environmental uncertainties. The resonant capture cross-section for thermal
neutrons by samarium-149 about two billion years ago ($z\simeq 0.15$) in the
Oklo natural nuclear reactor has created a samarium-149:samarium-147 ratio
at the reactor site that is depleted by the capture process $%
^{149}Sm+n\rightarrow ^{150}Sm+\gamma $ to an observed value of only about
0.02 compared to the value of about 0.9 found in normal samples of samarium.
The need for this capture resonance to be in place two billion years ago at
an energy level within about $90meV$ of its current value leads to very
strong bounds on all interaction coupling constants that contribute to the
energy level, as first noticed by Shlyakhter \cite{shly,jdb}. The latest
analyses by Fujii et al \cite{fuj} allow two solutions (one consistent with
no variation the other with a variation) because of the double-valued form
of the capture cross-section's response to small changes in the resonance
energy over the range of possible reactor temperatures: $\Delta \alpha
_{em}/\alpha _{em}\equiv -0.8\pm 1.0\times 10^{-8}$ or $\Delta \alpha
_{em}/\alpha _{em}\equiv 8.8\pm 0.7\times 10^{-8}.$ The latter possibility
does not include zero but might be excluded by further studies of other
reactor abundances. Subsequently, Lamoureax \cite{lam} has argued that a
better (non-Maxwellian) assumption about the thermal neutron spectrum in the
reactor leads to $6\sigma $ lower bound on the variation of $\Delta \alpha
_{em}/\alpha _{em}>4.5\times 10^{-8}$ at $z\simeq 0.15$.

Studies of the effects of varying a fine structure constant on the $\beta $%
-decay lifetime was first considered by Peebles and Dicke \cite{PD} as a
means of constraining allowed variations in $\alpha _{em}$ by studying the
ratio of rhenium to osmium in meteorites. The $\beta $-decay \ $%
_{75}^{187}Re\rightarrow _{76}^{187}Os+\bar{\nu}_{e}+e^{-}$ is very
sensitive to $\alpha _{em}$ and the analysis of new meteoritic data together
with new laboratory measurements of the decay rates of long-lived beta
isotopes has led to a time-averaged limit of $\Delta \alpha _{em}/\alpha
_{em}=8\pm 16\times 10^{-7}$ \cite{Olive} for a sample that spans the age of
the solar system ($z\leq 0.45$). Both the Oklo and meteoritic bounds are
complicated by the possibility of simultaneous variations of other constants
which contribute to the energy levels and decay rates; for reviews see refs. 
\cite{uzan, olive}. They also apply to environments within virialised
structures that do not take part in the Hubble expansion of the universe and
so it is not advisable to use them in conjunction with astronomical
information from quasars without a theory that links the values of $\alpha
_{em}$ in the two different environments that differ in density by a factor
of $O(10^{30}).$

In order to interpret these investigations it is essential to be in
possession of a self-consistent theory of the spacetime variation of $\alpha
_{em}$, analogous to the Brans-Dicke theory \cite{BD} for the variation of $%
G $, from which to derive further observational consequences of any inferred
variation. In the past, in the absence of any such theory, there has been a
tendency to produce limits on variations of couplings other than $\alpha
_{em}$ by simply `writing in' a time variation of the coupling into the
equations that hold when it is constant. Many of the quoted experimental
bounds on the variation of non-gravitational 'constants' that appear in the
literature have been arrived at in this way. Also questionable is the use of
laboratory bounds to limit possible variations of 'constants' on
extragalactic scales without any theory of the link between the two domains
of variation. Detailed discussions of this problem when $G$ and $\alpha $
vary have been made in refs. \cite{Mota:2004,Mota:2003}. A further problem
is the likelihood that if one coupling constant, like $\alpha _{em}$, varies
then others will vary also, especially in the presence of any pattern of
unification \cite{marc,drink}. Most deductions of bounds on constants assume
that only one constant is varying.

In order to deal with these problems of simultaneous variation and spatial
variation consistently, it is necessary to have self-consistent theories in
which `constants' vary through the dynamics of scalar fields, which
gravitate and conserve energy and momentum. Sandvik, Barrow and Magueijo
(SBM)\cite{Sandvik:2001} have extended Bekenstein's (B) \cite%
{Bekenstein:1982} original generalisation of Maxwell's theory to include
general relativity . This resulting BSBM theory provides a framework for
studying varying $\alpha _{em}$, and can be easily generalised to study the
simultaneous variation of $\alpha _{em}$ and $G$, as in ref. \cite{two}. A
number of detailed cosmological studies of the behaviour of this theory have
been made in refs. \cite%
{Barrow:2002a,Barrow:2002b,Barrow:2002c,Magueijo:2002,Barrow:2001,Sandvik:2001}%
. A further step has recently been taken by Kimberly and Magueijo \cite%
{Kimberly:2003} who have extended the BSBM theory to create a generalisation
of the Glashow-Salam-Weinberg theory to allow variation of the
electromagnetic and weak couplings. This allows the consequences of
electroweak unification to be investigated self consistently for the first
time. This approach has also been applied to the strong interaction alone by
Chamoun et al \cite{cham} and it is likely that a further step could be
taken to investigate the consequences of the simultaneous variation of all
gauge couplings in a grand unified theory. In this paper we determine the
quantum corrections to the Kimberly-Magueijo unfied electroweak models and
formulate the propagation equations for the two scalar fields that carry the
allowed time variations in the electromagnetic and weak couplings in a
universe populated by the particle spectrum of the standard model.

\section{The Simplest Model}

There has been strong interest in a class of scalar theories where $\alpha
_{em}=e^{2}/\hbar c$ can vary \cite%
{Bekenstein:1982,Barrow:2002a,Barrow:2002b,Barrow:2002c,Magueijo:2002,Barrow:2001,Sandvik:2001,bass,mart,beknew,dm}%
. These theories reduce to Maxwell's equations and general relativity in the
limiting case of no variation in the fine structure constant. In order to
evaluate the constraints introduced by a programme of unification, it is
important to extend these simple models to include electroweak and grand
unification. Kimberly and Magueijo \cite{Kimberly:2003} have proposed an
extensions to the Glashow-Salam-Weinberg (GSW) electroweak theory in which
the weak couplings can also vary in space and time and which reduces to the
standard GSW theory in the limit of no variation.

\subsection{A Single-Dilaton Theory}

The first model (KM-I) contains a single dilaton and allows both $\alpha
_{W}:=g_{W}^{2}$ and $\alpha _{em}$ to vary but their ratio, and hence the
mixing angle $\theta _{W}$, are true constants. In this model the
electroweak sector of the Standard Model is described by the following
lagrangian density: 
\begin{equation}
\mathcal{L}_{KM-I}=-\frac{1}{4}e^{-2\varphi }\left[ \mathbf{w}_{\mu \nu
}\cdot \mathbf{w}^{\mu \nu }+y_{\mu \nu }y^{\mu \nu }\right] +{\left( D_{\mu
}\mathbf{\Phi }\right) }^{\dagger }\left( D^{\mu }\mathbf{\Phi }\right) +%
\frac{\lambda }{4}\left( \mathbf{\Phi }^{\dagger }\mathbf{\Phi }%
-v^{2}\right) ^{2}-\frac{\omega }{2}\varphi _{,\mu }\varphi ^{,\mu },
\label{KMILagr}
\end{equation}%
where 
\begin{eqnarray}
&&\mathbf{w}_{\mu \nu }=2\mathbf{w}_{[\nu ,\mu ]}-\overline{g}_{W}\mathbf{w}%
_{\mu }\wedge \mathbf{w}_{\nu },  \label{wdef} \\
&&y_{\mu \nu }=2y_{[\nu ,\mu ]},  \label{ydef} \\
&&D_{\mu }\mathbf{\Phi }=\left( \partial _{\mu }-\frac{i}{2}\overline{g}_{W}%
\mathbf{t}\cdot \mathbf{w}_{\mu }-\frac{i}{2}\overline{g}_{Y}y_{\mu }\right) 
\mathbf{\Phi }  \label{hdef}
\end{eqnarray}%
and $\mathbf{\Phi }$ is the Higgs field. The dilaton field is $\varphi $ and 
$\omega $ is a dimensional parameter with units of $(mass)^{2}$. Since this
theory is perturbatively non-renormalisable, we would like $\omega =\mathcal{%
O}\left( M_{pl}^{2}\right) $ so that the dilaton enters at the same level as
gravity, and we are justified in ignoring any quantum fluctuations of the
dilaton field when quantising with respect to the other fields. The \textit{%
auxiliary} gauge fields, $\mathbf{w}_{\mu }$ and $y_{\mu }$, take values in
the adjoint representations of $\mathfrak{su}(2)$ and $\mathfrak{u}(1)$
respectively. They are not the \textit{physical} gauge fields, which will be
denoted by capital letters, but are related to them by the transformations 
\begin{eqnarray}
\bar{g}_{W}\mathbf{w}_{\mu } &=&g_{W}\mathbf{W}_{\mu },  \label{weqn} \\
\bar{g}_{Y}y_{\mu } &=&g_{Y}Y_{\mu },  \label{yeqn} \\
g_{W} &=&\bar{g}_{W}e^{\varphi },  \label{gweqn1} \\
g_{Y} &=&\bar{g}_{Y}e^{\varphi }.  \label{gyeqn1}
\end{eqnarray}

The distinction between the \textit{physical} and \textit{auxiliary} fields
will only become important at the quantum level (see section 2 below). When
written in terms of these \textit{auxiliary} gauge fields the covariant
derivatives which act upon matter species are independent of $\varphi $.
This makes it simpler to derive the classical field equations. The \textit{%
physical} gauge couplings, $g_{W}$ and $g_{Y}$, are dynamical whereas the 
\textit{auxiliary} couplings, $\bar{g}_{W}$ and $\bar{g}_{Y}$, are \textit{%
arbitrary} constants. At tree level the Fermi constant, $G_{F}$, and the
fermion masses do not vary, whereas the $W$ and $Z$ boson masses do.

We will also define the physical field strength tensors by 
\begin{eqnarray}
&&\mathbf{W}_{\mu \nu }:=\frac{\bar{g}_{W}}{g_{W}}\mathbf{w}_{\mu \nu }, \\
&&Y_{\mu \nu }:=\frac{\bar{g}_{Y}}{g_{Y}}y_{\mu \nu }.
\end{eqnarray}%
These field strengths reduce to the standard definitions of the weak and
hypercharge field strengths with gauge couplings $g_{W}$ and $g_{Y}$
respectively whenever the dilaton field $\varphi \ $is constant.

\subsection{A Two-Dilaton Theory}

In the second model (KM-II)proposed in ref. \cite{Kimberly:2003} a second
dilaton field is added. This results is the weak mixing angle, $\theta _{W}$%
, becoming a dynamical quantity. The lagrangian density of the electroweak
sector in this model is: 
\begin{equation}
\begin{split}
\mathcal{L}_{KM-II}=& -\frac{1}{4}e^{-2\varphi }\mathbf{w}_{\mu \nu }\cdot 
\mathbf{w}^{\mu \nu }-\frac{1}{4}e^{-2\chi }y_{\mu \nu }y^{\mu \nu }+{\left(
D_{\mu }\mathbf{\Phi }\right) }^{\dagger }\left( D^{\mu }\mathbf{\Phi }%
\right) +\frac{\lambda }{4}\left( \mathbf{\Phi }^{\dagger }\mathbf{\Phi }%
-v^{2}\right) ^{2} \\
& -\frac{\omega _{1}}{2}\varphi _{,\mu }\varphi ^{,\mu }-\frac{\omega _{2}}{2%
}\chi _{,\mu }\chi ^{,\mu }
\end{split}
\label{KMIILag}
\end{equation}%
The definitions (\ref{wdef})-(\ref{hdef}) still hold, as do relationships (%
\ref{weqn})-(\ref{yeqn}) but the \textit{physical} coupling constants are
now related to their \textit{auxiliary} values by the transformations 
\begin{eqnarray}
&&g_{W}=\bar{g}_{W}e^{\varphi }, \\
&&g_{Y}=\bar{g}_{Y}e^{\chi }, \\
&&\tan \theta _{W}=\left[ \frac{\bar{g}_{Y}}{\bar{g}_{W}}:=\tan \bar{\theta}%
_{W}\right] e^{\chi -\varphi }.
\end{eqnarray}%
The dilaton fields are $\varphi $ and $\chi $ and their respective
dimensionful scales are $\omega _{1}$ and $\omega _{2}$. As remarked above,
we would like $\omega _{i}\sim \mathcal{O}\left( M_{Pl}^{2}\right) $. In
accord with KM-I, $G_{F}$ and the fermion masses are constant at tree-level
whereas the boson masses are dynamical.

\subsection{Symmetry Breaking}

At low energies and temperatures, the most important feature of the GSW
electroweak model is that the $SU(2)_{L}\times U(1)_{Y}$ symmetry of the
lagrangian is spontaneously broken to $U(1)_{em}$ via the Higgs doublet, $%
\mathbf{\Phi }$, assuming a vacuum expectation value, $\mathbf{\Phi }_{0}$,
which minimises its potential. At tree-level this value is 
\begin{equation}
\mathbf{\Phi }_{0}=\left( {%
\begin{matrix}
0 \\ 
v%
\end{matrix}%
}\right) .
\end{equation}%
A perturbative expansion about this vacuum can be written as 
\begin{equation}
\mathbf{\Phi }=\left( {%
\begin{matrix}
0 \\ 
v+\frac{H\left( x^{\mu }\right) }{\sqrt{2}}%
\end{matrix}%
}\right) .
\end{equation}%
Expanding out the kinetic Higgs term gives: 
\begin{equation}
{\left( D_{\mu }\mathbf{\Phi }\right) }^{\dagger }\left( D^{\mu }\mathbf{%
\Phi }\right) =\frac{1}{2}H_{,\mu }H^{,\mu }+\frac{v^{2}(\overline{g}%
_{W})^{2}}{4}\left[ (\mathbf{w}_{\mu }^{1})^{2}+(\mathbf{w}_{\mu }^{2})^{2}%
\right] +\frac{v^{2}}{4}(\overline{g}_{W}\mathbf{w}_{\mu }^{3}-\overline{g}%
_{Y}y_{\mu })^{2}.  \label{hket}
\end{equation}%
When written in terms of the \textit{physical} gauge fields, the
broken-phase boson fields of both KM-I and KM-II are given by the usual
formulae: 
\begin{eqnarray}
&&W_{\mu }^{\pm }:=\frac{1}{\sqrt{2}}\left( \mathbf{W}_{\mu }^{1}\pm i%
\mathbf{W}_{\mu }^{2}\right) , \\
&&Z_{\mu }:=\frac{g_{W}\mathbf{W}_{\mu }^{3}-g_{Y}Y_{\mu }}{\sqrt{%
g_{W}^{2}+g_{Y}^{2}}}, \\
&&A_{\mu }:=\frac{g_{Y}\mathbf{W}_{\mu }^{3}+g_{W}Y_{\mu }}{\sqrt{%
g_{W}^{2}+g_{Y}^{2}}}.
\end{eqnarray}%
Hence, the tree-level boson masses and their dilaton field dependence can
read off as(\ref{hket}): 
\begin{eqnarray}
&&M_{W}=\frac{v}{\sqrt{2}}g_{W}, \\
&&M_{Z}=\frac{v}{\sqrt{2}}\sqrt{g_{W}^{2}+g_{Y}^{2}}, \\
&&M_{A}=0.
\end{eqnarray}

\subsection{Classical Field Equations}

In order to make and test the predictions of these theories we need to know
the field equations. In both KM-I and KM-II, the Einstein and Yang-Mills
equations are: 
\begin{eqnarray}
&&G_{\mu \nu }=8\pi G\left\{ \frac{\bar{g}_{W}^{2}}{g_{W}^{2}}T_{\mu \nu
}^{w}+\frac{\bar{g}_{Y}^{2}}{g_{Y}^{2}}T_{\mu \nu }^{y}+T_{\mu \nu }^{%
\mathbf{\phi }}+T_{\mu \nu }^{matter}+T_{\mu \nu }^{dilaton}\right\} , \\
&&\mathbf{D}^{\mu }\left( \frac{\bar{g}_{W}^{2}}{g_{W}^{2}}\mathbf{w}_{\mu
\nu }\right) =-\frac{\delta \mathcal{L}_{matter}}{\delta w^{\nu }}, \\
&&\partial ^{\mu }\left( \frac{\bar{g}_{Y}^{2}}{g_{Y}^{2}}\mathbf{y}_{\mu
\nu }\right) =-\frac{\delta \mathcal{L}_{matter}}{\delta y^{\nu }},
\end{eqnarray}%
where $T_{\mu \nu }^{w}$ and $T_{\mu \nu }^{y}$ are the standard Yang-Mills
energy-momentum tensors written in terms of the \textit{auxiliary} fields
and couplings. In KM-I the dilaton conservation equation is 
\begin{equation}
\square \varphi =-\frac{1}{2\omega }e^{-2\varphi }\left( \mathbf{w}_{\mu \nu
}\cdot \mathbf{w}^{\mu \nu }+y_{\mu \nu }y^{\mu \nu }\right) =-\frac{1}{%
2\omega }\left( \mathbf{W}_{\mu \nu }\cdot \mathbf{W}^{\mu \nu }+Y_{\mu \nu
}Y^{\mu \nu }\right) .  \label{d1}
\end{equation}%
and in KM-II we have conservation equations for the two fields: 
\begin{eqnarray}
&\square \varphi =&-\frac{1}{2\omega _{1}}e^{-2\varphi }\mathbf{w}_{\mu \nu
}\cdot \mathbf{w}^{\mu \nu }=-\frac{1}{2\omega _{1}}\mathbf{W}_{\mu \nu
}\cdot \mathbf{W}^{\mu \nu },  \label{d2} \\
&\square \chi =&-\frac{1}{2\omega _{2}}e^{-2\chi }y_{\mu \nu }\cdot y^{\mu
\nu }=-\frac{1}{2\omega _{2}}Y_{\mu \nu }Y^{\mu \nu }.  \label{d3}
\end{eqnarray}%
The conservation equations for other matter fields, like perfect fluids, are
unchanged. Although this system represents the full classical field
equations, their current form is not very useful. In order to do cosmology,
and make experimentally testable predictions, we need to understand how the
right-hand sides of (\ref{d1}, \ref{d2}, \ref{d3}) depend on macroscopic
quantities such as the background matter density and the value of the
dilaton field. It is this question that we address in the next section.

\section{The Dilaton-to-Matter Coupling}

Most tests which we might wish to apply to these, and other,
varying-constant theories require knowledge of how the dilaton fields will
evolve in the presence of some background matter density, $\rho $, and
pressure, $P$: 
\begin{equation*}
\rho :=\left\langle T_{0}{}^{0}{}^{(matter)}\right\rangle ,\quad P:=\frac{1}{%
3}\left\langle T_{i}{}^{i}{}^{(matter)}\right\rangle ,
\end{equation*}%
where $\left\langle \;\cdot \;\right\rangle $ denotes the quantum
expectation. In order to understand the dilaton evolution, we must therefore
evaluate the terms on the right-hand sides of (\ref{d1})-(\ref{d3}) under\
the$\left\langle \;\cdot \;\right\rangle $ operation. Those terms all
consist of terms quadratic in the Yang-Mills field strengths, i.e. $\mathbf{W%
}_{\mu \nu }\cdot \mathbf{W}^{\mu \nu }$ and $Y_{\mu \nu }Y^{\mu \nu }$,
henceforth we refer to them collectively as $F_{\mu \nu }F^{\mu \nu }$ or $%
F^{2}$ terms.

When we quantise this theory (leaving $\varphi $ as a classical field) we
must do so with respect to the \textit{physical} gauge fields rather than
the \textit{auxiliary} ones, since it is only the physical fields whose
kinetic terms possess the correct normalisation. This feature is important
if we want the renormalisation procedure to go through as usual when $%
\varphi =const$. It is for this reason that we have labelled the capitalised
fields as \textit{physical}.

\subsection{Derivation}

We need to understand how these $\left\langle F^{2}\right\rangle $ terms
depend, in a functional sense, on the dilaton fields and the background
matter density, $\rho $. For simplicity, we consider the case of a single
Dirac fermion, $\psi $, of mass $m$, coupled to a $U(1)$ gauge field $A_{\mu
}$ with Higgs scalar $\Phi $ and dilaton $\varphi $. The effective
lagrangian, including 't Hooft's gauge-fixing term, is: 
\begin{eqnarray}
&&\mathcal{L}=\mathcal{L}_{gauge,0}+\mathcal{L}_{fix}+\mathcal{L}_{\psi ,0}+%
\mathcal{L}_{int} \\
&&\mathcal{L}_{gauge,0}=-\tfrac{1}{4}F^{\mu \nu }F_{\mu \nu }+\left\Vert
\left( \partial _{\mu }+i\tfrac{M_{A}}{v}A_{\mu }\right) \Phi \right\Vert
^{2}-\tfrac{\lambda }{4}\left( \Phi ^{\dagger }\Phi -\tfrac{v^{2}}{2}\right)
^{2} \\
&&\mathcal{L}_{fix}=-\frac{1}{2\xi }\left( \partial ^{\mu }A_{\mu }+\xi M_{A}%
\mathrm{Im}\left( \Phi \right) \right) ^{2} \\
&&\mathcal{L}_{\psi ,0}=\bar{\psi}\left( i\gamma ^{\mu }\partial _{\mu
}-m\right) \psi \\
&&\mathcal{L}_{int}=-e\left( \varphi \right) \bar{\psi}\gamma ^{\mu }\psi
A_{\mu }
\end{eqnarray}%
The ghost fields have been excluded since in this gauge they decouple from
the abelian sector. By taking a spacetime region $\mathcal{R}$, of volume $V$
and temporal extent $T$, that is large compared to the scale of quantum
fluctuations of the matter and gauge fields, and yet small when compared to
the scale over which the dilaton field varies, we can quantise this theory
inside $\mathcal{R}$ by taking $\varphi =const$ and defining the partition
function in the usual manner. For energies far below the Higgs mass, $m_{H}$%
, we can ignore quantum Higgs fluctuations in the first approximation. Doing
this and integrating out the gauge fields we determine that 
\begin{equation}
\begin{split}
-\tfrac{1}{4}\left\langle F_{\mu \nu }F^{\mu \nu }\right\rangle =& \frac{%
e^{2}(\varphi )}{2VT}\int \int \int \mathrm{d}^{4}x\mathrm{d}^{4}y\mathrm{d}%
^{4}z\Big <\partial _{x}^{2}\mathcal{D}_{\mu \nu }^{gauge}(x-y)\left( \bar{%
\psi}\gamma ^{\nu }\psi \right) (y)\mathcal{D}^{\mu }{}_{\rho
}{}^{gauge}(x-z)\left( \bar{\psi}\gamma ^{\rho }\psi \right) (z) \\
& +\partial ^{\mu }{}_{x}\mathcal{D}_{\mu \nu }^{gauge}(x-y)\left( \bar{\psi}%
\gamma ^{\nu }\psi \right) (y)\partial ^{\tau }{}_{x}\mathcal{D}_{\tau \rho
}^{gauge}(x-z)\left( \bar{\psi}\gamma ^{\rho }\psi \right) (z)\Big >_{\Psi },
\end{split}
\label{F2Eqn}
\end{equation}%
where the gauge-field propagator is given by: 
\begin{equation*}
\mathcal{D}_{\mu \nu }^{gauge}(x)=\int \frac{\mathrm{d}^{4}k}{\left( 2\pi
\right) ^{4}}ie^{ikx}\left[ -g^{\mu \nu }+\frac{\left( 1-\xi \right) k^{\mu
}k^{\nu }}{k^{2}-\xi M_{A}^{2}}\right] \frac{1}{k^{2}-M_{A}^{2}}
\end{equation*}%
The quantum expectation operator, $\left\langle \cdot \right\rangle _{\Psi }$%
, used in equation (\ref{F2Eqn}), is defined thus: 
\begin{equation}
\left\langle A\left( \psi ,\bar{\psi}\right) \right\rangle _{\Psi }:=\frac{%
\int \left[ \mathrm{d}\bar{\psi}\right] \left[ \mathrm{d}\psi \right]
A\left( \psi ,\bar{\psi}\right) \exp {\left( -i\frac{e^{2}}{2}\int \int 
\mathrm{d}^{4}x\mathrm{d}^{4}yJ^{\nu }\left( x\right) \mathcal{D}%
^{gauge}\left( x-y\right) J^{\nu }\left( y\right) \right) }e^{i\int \mathrm{d%
}^{4}x\mathcal{L}_{\psi ,0}}}{\int \left[ \mathrm{d}\bar{\psi}\right] \left[ 
\mathrm{d}\psi \right] \exp {\left( -i\frac{e^{2}}{2}\int \int \mathrm{d}%
^{4}x\mathrm{d}^{4}yJ^{\nu }\left( x\right) \mathcal{D}^{gauge}\left(
x-y\right) J^{\nu }\left( y\right) \right) }e^{i\int \mathrm{d}^{4}x\mathcal{%
L}_{\psi ,0}}}  \label{avgeqn}
\end{equation}%
The $\int \left[ \mathrm{d}\psi \right] $ represents a functional integral
with respect to the $\psi $-field, and we have defined $J^{\mu }(x):=\bar{%
\psi}(x)\gamma ^{\mu }\psi (x)$. The term with an exponent that is quadratic
in the $J$'s in equation (\ref{avgeqn}) encodes the self-energy corrections
to the fermion and gauge boson propagators. By writing both the fermion and
gauge boson propagators as full propagators we are justified in dropping the
aforementioned exponential term because its effect is of sub-leading order.
Finally then, by writing this expression in momentum space we reduce it to
the more transparent and manifestly gauge invariant form: 
\begin{equation}
-\tfrac{1}{4}\left\langle F_{\mu \nu }F^{\mu \nu }\right\rangle =\frac{%
e^{2}\left( \varphi \right) }{2VT}\int \frac{\mathrm{d}^{4}p}{\left( 2\pi
\right) ^{4}}p^{2}\left[ \left( p^{2}-M_{A}^{2}(\varphi )\right) g^{\mu \nu
}+\Pi _{A}^{\mu \nu }\left( p,e^{2}\left( \varphi \right) \right] \right)
^{-2}\left\langle \tilde{J}^{\mu }\left( p\right) \tilde{J}^{\nu }\left(
-p\right) \right\rangle _{\Psi \ast },  \label{mtmeqn}
\end{equation}%
where $\tilde{J}_{\mu }$ is the momentum space representation of $J^{\mu }$
and $\Pi _{A}^{\mu \nu }$ is the vacuum polarisation of the gauge boson. The
expectation here is defined with respect to the partition function 
\begin{equation*}
Z_{\Psi \ast }:=\int \left[ \mathrm{d}\bar{\psi}\right] \left[ \mathrm{d}%
\psi \right] \exp \left( i\int \mathrm{d}^{4}x\mathrm{d}^{4}y\bar{\psi}%
(x)K(x-y)\psi (y)\right)
\end{equation*}%
with $K(x-y)$ denoting the inverse of the full fermion propagator.

\subsection{Interpretation}

By Wick's theorem it is apparent that the remaining quantum expectation in (%
\ref{mtmeqn}) contains two distinct contributions. The first contribution
will always involve boson exchange between two fermionic particles and so is
proportional to $\left( \rho -3P\right) ^{2}/m^{2}$ to leading order. The
second contribution results from a single fermionic particle admitting and
reabsorbing a gauge boson (i.e. the process that results in the fermion's
self-energy). This second term will clearly be proportional to $m^{2}\left(
\rho -3P\right) $ to leading order. In almost all cases of experimental
interest $\left( \rho -3P\right) /m^{4}\ll 1$. Only in objects whose density
approaches that of nuclear matter does it fail to hold and this naïve
quantisation procedure is not suitable for dealing with such high-density
backgrounds. For this reason we drop the contribution due to photon exchange.

When the perturbation theory holds it is appropriate to expand (\ref{mtmeqn}%
) as a series in the gauge coupling, $e^{2}\left( \varphi \right) $, giving 
\begin{equation}
-\tfrac{1}{4}\left\langle F_{\mu \nu }F^{\mu \nu }\right\rangle =e^{2}\zeta
\left( \varphi \right) \left( \rho -3P\right) \Big(1+\mathcal{O}\left(
e^{2}\ln e\right) \Big)\left( 1+\mathcal{O}\left( \frac{\rho }{m^{4}}\right)
\right)  \label{feff}
\end{equation}%
where, for this model, $\zeta $ is a defined by: 
\begin{equation}
\zeta \left( \varphi \right) \left( \rho -3P\right) =\int \int \frac{\mathrm{%
d}^{4}p}{\left( 2\pi \right) ^{4}}\frac{\mathrm{d}^{4}q}{\left( 2\pi \right)
^{2}}\frac{p^{2}}{\left( p^{2}-M_{A}^{2}\left( \varphi \right) \right) ^{2}}%
\mathrm{tr}\left( \gamma ^{\nu }\frac{q\cdot \gamma +m}{q^{2}-m}\gamma _{\nu
}\frac{(p-q)\cdot \gamma +m}{(p-q)^{2}-m}\right)  \label{zeta}
\end{equation}

When inter-fermion strong interactions are introduced, confinement may
occur. If this happens the trace factor in (\ref{zeta}) will take on a more
complicated form. At leading order, however, it will remain independent of $%
\varphi $ so long as $\Lambda _{QCD}$ is. At densities much lower than the
nucleon mass, the right-hand side will be proportional to the hadronic
energy density.

\subsection{The Unbroken Symmetry Case: $M_{A}=0$}

It is helpful to have a physical interpretation of the $\zeta $ parameter.
When dynamical symmetry breaking does not occur (and so we have $M_{A}^{2}=0$%
) the right-hand side of $\left( \ref{zeta}\right) $ is proportional to the
1-loop fermion field self-energy resulting from its interaction with the $%
A_{\mu }$ gauge field. Defining this self-energy to be $\delta m^{2}\left(
\varphi \right) $ we have: 
\begin{equation}
\zeta =\frac{\delta m^{2}\left( \varphi \right) }{e^{2}\left( \varphi
\right) m^{2}}=\frac{\delta \bar{m}^{2}}{\bar{e}^{2}m^{2}}=const\text{.}
\label{zetau}
\end{equation}%
In this case $\zeta $ is $\varphi $-\textit{independent} at leading order. $%
\delta \bar{m}$ is defined as the electromagnetic mass correction when the
electric charge is some value $\bar{e}=const$.

\subsection{The Broken Symmetry Case: $M_{A}\neq 0$}

When $M_{A}^{2}\neq 0$ the above interpretation of $\zeta $ in terms of the
self-energy mass-correction will, in general, fail. The correct physical
interpretation will depend heavily on the size of $\frac{m}{M_{A}\left(
\varphi \right) }$. We consider the three cases:

\begin{itemize}
\item $\frac{m}{M_{A}}\ll 1$. The dominant contribution to the $\zeta $
integral will come from momenta $p^{2}\gg M_{A}^{2}$, and so we can ignore $%
M_{A}$ at leading order and interpret $\zeta $ just as we did in the $%
M_{A}=0 $ case. Hence, $\zeta $ is dilaton independent to leading order.

\item $\frac{m}{M_{A}}\gg 1$. The dominant contribution to the (\ref{zeta})
will come from momenta $p^{2}\approx m^{2}$. Hence 
\begin{equation}
\zeta \left( \varphi \right) =\left( \frac{m}{M_{A}\left( \varphi \right) }%
\right) ^{4}\zeta _{0},
\end{equation}%
where $\zeta _{0}$ is of the order the value of $\zeta $ which we would have
had if $M_{A}=0$. In this case $\zeta $ will vary with $\varphi $ like $%
\frac{1}{M_{A}^{4}}$.

\item $\frac{m}{M_{A}}=\mathcal{O}\left( 1\right) $. This is by far the most
difficult case to analyse. In general ,$\zeta $ will have a leading-order
dependence on the dilaton field through $M_{A}\left( \varphi \right) $, but
the precise form $\zeta \left( \varphi \right) $ will depend on the nature
of the trace term in equation (\ref{zeta}). This in turn rests on the
precise details of the microscopic physical model for the matter fields in
question. This chain of complications means that a general prescription for
the dilaton dependence of $\zeta $, in this case, is not possible.
\end{itemize}

\section{Applications to the Kimberly-Magueijo Models}

\subsection{Ultra-Relativistic Matter}

At the very high energies required to restore the broken gauge symmetry in
these theories, most matter species will be ultra-relativistic and we will
assume this to be the case for all species. We will also assume that there
is a discrete spectrum of spin states. Matter will therefore behave like
black-body radiation with $\rho =3P$. It is clear from equation (\ref{feff})
that these ultra-relativistic species do not contribute to the $\left\langle
F^{2}\right\rangle $ terms which source the dilaton fields' evolution.

The only uncharged fundamental field is the neutrino, which we will assume
to be so light that it remains relativistic at experimental and cosmological
temperatures. Such neutrinos make only a very small contribution to the
dilaton source terms. The corollary of this is that amongst the known
fundamental matter species, we are justified in assuming that only those
with $\ $non-zero charge contribute to the right-hand sides of equations (%
\ref{d1})-(\ref{d3}).

The see-saw mechanism for neutrino mass-generation results in three light
neutrinos, masses $m_{\nu }^{(i)}$ (corresponding to the currently observed
particles) and three very heavy neutrinos, masses $M_{N}^{(i)}\approx $ a
few tens of GeV. The light neutrinos are formed mostly from the weakly
interacting left-handed components, whilst the heavy neutrinos are primarily
composed of the weak singlet right-handed particles. Such heavy neutrinos
would therefore interact with the weak bosons much more weakly that other
massive particle species. To zeroth order in $m_{\nu }^{(i)}/M_{N}^{(i)}\ll
1 $, we can assume these heavy neutrinos to be non-interacting with either
electromagnetic or weakly interacting particles. Their contribution to the $%
\left\langle F^{2}\right\rangle $ terms will be negligible compared to that
of the other matter species.

\subsection{The Top Quark}

With a mass of about 180 GeV, the top quark is the only fermionic species
present in the Standard Model that does not fall into the $\frac{m}{M_{W}}%
\ll 1$ category. Whilst in principle its contribution to the dilaton terms
could be calculated, it would be a difficult procedure which would require
accounting for gauge boson self-interactions, Higgs boson fluctuations, and
non-perturbative effects - as well as having to do QCD calculations.
However, the results would actually have very little bearing upon most
cosmological tests of this theory, since the background energy density of
top quarks is entirely negligible when compared to that of all the other
forms of matter.

\subsection{`Light' Charged Matter}

We have just argued that, amongst the Standard Model particle species, the
only non-negligible contributions to the dilaton source terms will come from
particles with mass $m$ and charge $Q\neq 0$, which are light compared to $%
M_{W}$. We must also require the particles to be relativistic, allowing us
to restrict to the low-energy broken symmetric phase of the KM electroweak
theories. This is appropriate for energies well below the $100GeV$ level. At
these energies, perturbation theory will be appropriate and non-abelian
effects will be of sub-leading order. In this case then, the results of
Section II should be valid.

Writing the $\left\langle \mathbf{W}_{\mu \nu }\cdot \mathbf{W}^{\mu \nu
}\right\rangle $ and $\left\langle Y_{\mu \nu }\cdot Y^{\mu \nu
}\right\rangle $ quantities in terms of the low-energy physical fields we
find: 
\begin{equation}
\begin{split}
\left\langle -\tfrac{1}{4}\mathbf{W}_{\mu \nu }\cdot \mathbf{W}^{\mu \nu
}\right\rangle =& \Big[\left\langle -\tfrac{1}{2}F_{W\pm }^{\dagger }\cdot
F_{W\pm }\right\rangle +\sin ^{2}\theta _{W}\left\langle -\tfrac{1}{4}%
F_{em}^{2}\right\rangle +\cos ^{2}\theta _{W}\left\langle -\tfrac{1}{4}%
F_{Z}^{2}\right\rangle +\sin \theta _{W}\cos \theta _{W}\left\langle -\tfrac{%
1}{2}F_{em}\cdot F_{Z}\right\rangle \Big] \\
& \cdot \Big[1+\mathcal{O}\left( \partial _{\mu }\theta _{W}\right) +%
\mathcal{O}\left( g_{W},g_{Y}\right) \Big]
\end{split}
\label{Wbreak}
\end{equation}%
\begin{equation}
\left\langle -\tfrac{1}{4}Y_{\mu \nu }\cdot Y^{\mu \nu }\right\rangle =\Big[%
\cos ^{2}\theta _{W}\left\langle -\tfrac{1}{4}F_{em}^{2}\right\rangle +\sin
^{2}\theta _{W}\left\langle -\tfrac{1}{4}F_{Z}^{2}\right\rangle -\sin \theta
_{W}\cos \theta _{W}\left\langle -\tfrac{1}{2}F_{em}\cdot F_{Z}\right\rangle %
\Big]\cdot \Big[1+\mathcal{O}\left( \partial _{\mu }\theta _{W}\right) \Big]
\label{Ybreak}
\end{equation}

The $\mathcal{O}\left(\partial_{\mu}\theta_{W}\right)$ terms produce only
negligible corrections. The $\mathcal{O}\left(g_W,g_Y\right)$ symbol
represents the additional terms that arise from gauge boson self
interactions; these only contribute at sub-leading order.

Thus for a non-relativistic particle species, $\psi _{i}$, of mass $m_{i}\ll
M_{W}$, charge $Q_{i}\neq 0$, and weak isospin $t_{3i}$ we see that only $%
\left\langle -\tfrac{1}{4}F_{em}^{2}\right\rangle $ will contribute to (\ref%
{Wbreak}, \ref{Ybreak}) at leading order. The $\left\langle -\tfrac{1}{4}%
F_{Z}^{2}\right\rangle $, $\left\langle -\tfrac{1}{4}\left\vert F_{W\pm
}^{2}\right\vert \right\rangle $ and $\left\langle -\tfrac{1}{2}F_{em}\cdot
F_{Z}\right\rangle $ terms are suppressed by relative factors of order of $%
\frac{g_{W}^{2}}{Q_{i}^{2}e^{2}\cos ^{2}\theta _{W}}\frac{m_{i}^{4}}{%
M_{Z}^{4}}\left( t_{3i}-2Q\sin ^{2}\theta _{W}\right) ^{2}$, $\frac{g_{W}^{2}%
}{Q_{i}^{2}e^{2}}\frac{m_{i}^{4}}{M_{W}^{4}}t_{3i}^{2}$ and $\frac{g_{W}}{%
Q_{i}e\cos \theta _{W}}\frac{m_{i}^{2}}{M_{Z}^{2}}\left( t_{3i}-Q_{i}\sin
^{2}\theta _{W}\right) $ respectively. The leading-order contributions of
such a `light' matter species to (\ref{Wbreak}) and (\ref{Ybreak}) therefore
reduce to 
\begin{equation}
\begin{split}
& \left\langle -\tfrac{1}{4}\mathbf{W}_{\mu \nu }\cdot \mathbf{W}^{\mu \nu
}\right\rangle _{i}=\sin ^{2}\theta _{W}\frac{e^{2}}{\bar{e}^{2}}\frac{%
\delta \bar{m}_{i}^{2}}{m_{i}^{2}}\rho _{i}, \\
& \left\langle -\tfrac{1}{4}Y_{\mu \nu }Y^{\mu \nu }\right\rangle _{i}=\cos
^{2}\theta _{W}\frac{e^{2}}{\bar{e}^{2}}\frac{\delta \bar{m}_{i}^{2}}{%
m_{i}^{2}}\rho _{i},
\end{split}
\label{locontrib}
\end{equation}%
where $\delta \bar{m}$ is defined as the electromagnetic mass correction
when the electric charge is some value $\bar{e}=const$.

\subsection{Dark Matter}

There is a great deal of evidence from cosmological and astronomical data
for the existence of dark matter which contributes about 27\% of the
gravitating mass density of the universe. Such matter must be non-baryonic,
electromagnetically non-interacting and non-relativistic ('cold'). It
appears that the Standard Model lacks any suitable dark matter candidates.
Weakly Interacting Massive Particles (WIMPS), such as the lightest putative
supersymmetric partners in MSSM, are one possible class of candidates for
dark matter. The masses of these particles tend to be of the order of a few
tens of GeVs and so fall into the $m\approx M_{W}$ category. Locally, if
dynamically virialised, they will have $keV$ energies which may allow them
to be detected in underground nuclear recoil experiments. They are
necessarily uncharged ( $Q\neq 0$) but they can however interact weakly and
as such contribute to the $\left\langle -\tfrac{1}{4}F_{Z}^{2}\right\rangle $
and $\left\langle -\tfrac{1}{4}\left\vert F_{W\pm }^{2}\right\vert
\right\rangle $ terms. Thus the leading-order contribution from these WIMPs
to (\ref{Wbreak}, \ref{Ybreak}) would be given by: 
\begin{equation}
\begin{split}
& \left\langle -\tfrac{1}{4}\mathbf{W}_{\mu \nu }\cdot \mathbf{W}^{\mu \nu
}\right\rangle _{wimp}=\frac{g_{W}^{2}}{\bar{g}_{W}^{2}}\left[ \mathcal{F}%
_{W}\left( \frac{m_{wimp}^{2}}{M_{W}^{2}}\right) +\mathcal{F}_{Z}\left( 
\frac{m_{wimp^{2}}}{M_{Z}^{2}}\right) \right] \rho _{wimp}, \\
& \left\langle -\tfrac{1}{4}Y_{\mu \nu }Y^{\mu \nu }\right\rangle _{wimp}=%
\frac{g_{W}^{2}}{\bar{g}_{W}^{2}}\tan ^{2}\theta _{W}\mathcal{F}_{Z}\left( 
\frac{m_{wimp}^{2}}{M_{Z}^{2}}\right) \rho _{wimp},
\end{split}
\label{wimpcontrib}
\end{equation}%
where we have defined $\mathcal{F}_{W}$ and $\mathcal{F}_{Z}$ to be
WIMP-model dependent `structure' functions. These encode precisely how the
WIMP's $\zeta $ parameters depend on the gauge boson masses. We expect $%
\left\vert \mathcal{F}_{W}\right\vert $ and $\left\vert \mathcal{F}%
_{Z}\right\vert $ to be $\ll 1$. Not much more can be said about their
structure as functions of the dilaton field without the aid of a microscopic
model for dark matter. We shall therefore leave this discussion for a
separate work. It should be noted also that, if $\mathcal{F}_{W}$ and $%
\mathcal{F}_{Z}$ are no more than an order of magnitude or so smaller than
the $\zeta _{em}$ parameter of baryonic matter, then WIMP-based dark matter
could well be the dominant factor in the cosmological evolution of the
dilaton fields because the cosmological dark matter density in WIMPs is an
order of magnitude greater than that of baryonic matter.

\section{Discussion}

In this paper we have derived the form of the effective dilaton field
equations for an electroweak theory with varying couplings in the presence
of a background matter density. It is this form that is of most use for
experimental, observational and cosmological tests of this theory. We took
particular care in the identification of the \textit{physical} gauge fields.
In previous work on BSBM theory, the dilaton dependence of the $F^{2}$
source term was incorrectly determined as a result of the auxiliary fields
mistakenly taken to be the physically propagating modes. This error resulted
in the statement: 
\begin{equation}
\frac{\bar{e}^{2}}{e^{2}}\left\langle -\frac{1}{4}f_{\mu \nu }f^{\mu \nu
}\right\rangle =\frac{\bar{e}^{2}}{e^{2}}\frac{\delta \bar{m}^{2}}{m^{2}}%
=e^{-2\varphi }\zeta \rho  \label{bsbm}
\end{equation}%
Where $f_{\mu \nu }=2\partial _{\lbrack \mu }a_{\nu ]}$ was the auxiliary
field strength, and $a_{\mu }$ the auxiliary photon field. $\delta \bar{m}$
is defined as the electromagnetic mass correction when the electric charge
is some value $\bar{e}=const$. This lead to the conclusion that the
leading-order dilaton field dependence went like: $\frac{\bar{e}^{2}}{e^{2}}%
=e^{-2\varphi }$. However, it is clear from equations (\ref{zeta})-(\ref%
{zetau}) that in reality the leading-order dilaton dependence should be $%
\frac{e^{2}}{\bar{e}^{2}}=e^{+2\varphi }$ and equation (\ref{bsbm}) should
read: 
\begin{equation}
\left\langle -\frac{\bar{e}^{2}}{e^{2}}\frac{1}{4}f_{\mu \nu }f^{\mu \nu
}\right\rangle =\frac{e^{2}}{\bar{e}^{2}}\frac{\delta \bar{m}^{2}}{m^{2}}%
\rho =e^{2\varphi }\zeta \rho  \label{bsbmright}
\end{equation}

Indeed equation (\ref{bsbm}) is problematic because, in the limit of zero
electric charge, matter decouples from the photon, and so cannot possibility
contribute to the $F_{em}^{2}$ term. This situation corresponds to $\varphi
\rightarrow -\infty $. However, in this limit, the $F_{em}^{2}$ term as
given by the right-hand side of equation (\ref{bsbm}) grows infinity large.
Equation (\ref{bsbmright}) shows the correct behaviour, and vanishes, as it
should, it this limit.

We have also seen that if the gauge bosons become massive then, for
particles which are much lighter than the gauge boson in question, the
leading-order dilaton dependence of the $F^{2}$ term changes from $g^{2}$ ($%
g $ is the physical gauge coupling) to $\frac{g^{2}}{M_{gauge}^{4}}\sim 
\frac{1}{g^{2}}$. For particles that are much heavier than the gauge boson
in question, provided perturbation theory is still valid at energies of the
order of the particle's mass, the leading-order dilaton dependence remains
as $g^{2}$. We also briefly discussed the complication of $%
m_{particle}\approx M_{gauge}$, whereby the leading-order dilaton dependence
will be highly susceptible to the details of the matter model in question,
and noted that this effect might be important in cosmology if dark matter is
weakly interacting.

In all of this analysis we assumed both that perturbation theory holds and
that any non-abelian effects are negligible. Whilst this is true for
electroweak theory at energies well below the Higgs boson mass, it will not
be true for QCD. If we were to construct a BSBM-like varying $\alpha
_{strong}=g_{strong}^{2}$ (or equivalently, varying $\Lambda _{QCD}$)
theory, we would expect the leading-order dilaton dependence of the $%
F_{strong}^{2}$ term to come from a complicated function of $g_{strong}$.
Evaluating this function would, at the very least, require us to be able to
predict quark and nucleon masses accurately via a QCD calculation which is
not yet possible. In the absence of such calculations, varying $\Lambda
_{QCD}$ theories will be difficult to test accurately or make use of in the
early universe except at the very highest energies. Fortunately, we do not
encounter these problems in the electroweak KM theories.

We conclude with a statement of the effective dilaton field equations we
have derived. In the KM-I theory these are: 
\begin{equation}
\square \varphi =\frac{2}{\omega }e^{2\varphi }\sum_{i}\zeta _{i}\rho _{i}+%
\frac{2}{\omega }e^{2\varphi }\left[ \mathcal{F}_{W}\left( \frac{m_{wimp}^{2}%
}{M_{W}\left( \varphi \right) ^{2}}\right) +\sec ^{2}\theta _{W}\mathcal{F}%
_{Z}\left( \frac{m_{wimp}^{2}}{M_{Z}\left( \varphi \right) ^{2}}\right) %
\right] \rho _{wimp,}
\end{equation}%
where the sum is over all charged matter species (with $m\ll M_{W}$). The
first term is identical to the properly evaluated source term in BSBM. Hence
the effective KM-I theory differs (at low energies and densities) from BSBM
only in the putative WIMP matter contribution. If we transform $\varphi
\rightarrow -\varphi $ then we can read off the explicit correspondence with
the the studies in refs. \cite%
{Mota:2004,Mota:2003,Barrow:2002a,Barrow:2002b,Barrow:2002c,Magueijo:2002,Barrow:2001,Sandvik:2001}%
. When $\zeta /\omega <0$ we obtain slow logarithmic growth of the fine
structure 'constant' during the dust dominated era, as $\varphi \propto \ln
[\ln (t+t_{0})],$ $t_{0}$ constant but constant-$\varphi $ behaviour during
the radiation and dark-energy dominated eras. If $\zeta /\omega >0$ then the
solutions predict a much stronger evolution of $\alpha $ with time that is
difficult to reconcile with the observational constraints. The sign of $%
\zeta /\omega $ is controlled by the sign of $\zeta \in \lbrack -1,1]$ and
of $\omega $. Positive $\zeta $ corresponding to 'normal' matter dominated
by the electrostatic contributions, and negative $\zeta $ corresponding to
matter (like superconducting cosmic strings, which have $\zeta =-1$) that is
dominated by magnetic energy. Positive $\omega $ corresponds to a positive
kinetic contribution to the energy by the $\varphi $ field while negative $%
\omega $ indicates that it is a ghost field, as in ref. \cite{barrow:2004}.

In the KM-II theory the effective dilaton equations are: 
\begin{eqnarray}
&\square \varphi =&\frac{2}{\omega _{1}}\sin ^{2}\theta _{W}\frac{\alpha
_{em}\left( \varphi ,\chi \right) }{\bar{\alpha}_{em}}\sum_{i}\zeta _{i}\rho
_{i}+\frac{2}{\omega _{1}}e^{2\varphi }\left[ \mathcal{F}_{W}\left( \frac{%
m_{wimp}^{2}}{M_{W}\left( \varphi \right) ^{2}}\right) +\mathcal{F}%
_{Z}\left( \frac{m_{wimp}^{2}}{M_{Z}\left( \varphi ,\chi \right) ^{2}}%
\right) \right] \rho _{wimp} \\
&\square \chi =&\frac{2}{\omega _{2}}\cos ^{2}\theta _{W}\left( \varphi
,\chi \right) \frac{\alpha _{em}\left( \varphi ,\chi \right) }{\bar{\alpha}%
_{em}}\sum_{i}\zeta _{i}\rho _{i}+\frac{2}{\omega _{2}}e^{2\chi }\tan ^{2}%
\bar{\theta}_{W}\mathcal{F}_{Z}\left( \frac{m_{wimp}^{2}}{M_{Z}\left(
\varphi ,\chi \right) ^{2}}\right) \rho _{wimp}
\end{eqnarray}%
A similar menu of possibilities exists for the sign of the leading term on
the right-hand side of the $\square \varphi $ and $\square \chi $ equations
as was the case for the KM-I theory discussed above. In the absence of the
permitted dark matter contributions, this two-dilaton theory will only
reduce to BSBM when $\omega _{2}\sin ^{2}\theta _{W}=\omega _{1}\cos
^{2}\theta _{W}$. In all other cases $\theta _{W}$ will vary and lead to an
evolution of $\alpha _{em}$ that is different from that of BSBM theory.
Further cosmological consequences of these results will be explored
elsewhere.

\textbf{Acknowledgements} DS is supported by a PPARC studentship. We would
like to thank J.K. Webb, D. Kimberly and J. Magueijo for helpful discussions.


\begin{thebibliography}{99}
\bibitem{Webb:2001} J. K. Webb, M.T. Murphy, V. Flambaum, V. Dzuba, J.D.
Barrow, C. Churchill, J. Prochaska, and A. Wolfe, Phys. Rev. Lett. \textbf{87%
}, 091301 (2001).

\bibitem{Murphy:2001} M.T. Murphy \textit{et al.}, Mon. Not. R. astron. Soc.%
\textit{.} \textbf{327}, 1208 (2001).

\bibitem{Webb:1999} J. K. Webb, V.V. Flambaum, C.W. Churchill, M.J.
Drinkwater and J. D. Barrow, Phys. Rev. Lett. \textbf{82,} 884 (1999).

\bibitem{lab1} H. Marion, \textit{et al.}, Phys. Rev. Lett.\textbf{\ 90},
150801 (2003).

\bibitem{lab2} S. Bize, \textit{et al.}, Phys. Rev. Lett. \textbf{90},
150802 (2003).

\bibitem{lab3} M. Fischer, \textit{et al.,} Phys. Rev. Lett. \textbf{92},
230802 (2004).

\bibitem{lab4} E. Peik \textit{et al.,} physics/0402132.

\bibitem{chand1} H. Chand \textit{et al.}, Astron. Astrophys. \textbf{417},
853 (2004).

\bibitem{chand2} R. Srianand \textit{et al.}, Phys. Rev. Lett. \textbf{92},
121302 (2004).

\bibitem{sdss} J. Bahcall, C.L. Steinhardt, and D. Schlegel, Astrophys. J.%
\textbf{\ 600}, 520 (2004).

\bibitem{qu} R. Quast, D. Reimers and S.A Levashov, astro-ph/0311280.

\bibitem{lev} S.A. Levashov, \textit{et al.,} astro-ph/0408188.

\bibitem{darl} J. Darling, Phys. Rev. Lett. \textbf{91}, 011301 (2003).

\bibitem{oh} J. Darling, astro-ph/0405240.

\bibitem{drink} M.J. Drinkwater, J.K. Webb, J.D. Barrow and V.V. Flambaum,
Mon. Not. R. Astron. Soc. \textbf{295}, 457 (1998).

\bibitem{tz} P. Tzanavaris, J.K. Webb, M.T. Murphy, V.V. Flambaum, and S.J.
Curran, astro-ph/0412649.

\bibitem{cmb} P.P. Avelino \textit{et al.}, Phys. Rev. D \textbf{62}, 123508
(2000) and Phys. Rev. D \textbf{64}, 103505 (2001); R.A. Battye, R.
Crittenden and J. Weller, Phys. Rev. D \textbf{63,} 0453505 (2001); S.
Landau, D.D. Harari, and M. Zaldarriaga, Phys. Rev. D \textbf{63}, 083505
(2001); C. Martins \textit{et al.}, Phys. Lett. B \textbf{585}, 29 (2004);
G. Rocha \textit{et al}., N. Astron. Rev. \textbf{47}, 863 (2003); K.
Sigurdson, \textit{et al.}, Phys. Rev. D \textbf{68}, 103509 (2003).

\bibitem{bbn} J.D. Barrow, Phys. Rev. D \textbf{35}, 1805 (1987); B.A.
Campbell, and K.A Olive, Phys. Lett. B \textbf{345}, 429 (1995); R.H. Cyburt,%
\textit{\ et al.}, astro-ph/0408033.

\bibitem{shly} A.I. Shlyakhter, Nature \textbf{264}, 340 (1976).

\bibitem{jdb} J.D. Barrow, \textit{The Constants of Nature: from alpha to
omega}, (Vintage, London, 2002).

\bibitem{fuj} Y. Fujii, \textit{et al.}, Nucl. Phys. B \textbf{573}, 377
(2000).

\bibitem{lam} S.K. Lamoureaux, Phys. Rev. D \textbf{69}, 121701 (2004).

\bibitem{PD} P.J. Peebles and Dicke, R.H., Phys. Rev. \textbf{128}, 2006
(1962).

\bibitem{Olive} K.A. Olive \textit{et al}., Phys. Rev. D \textbf{66}, 045022
(2000).

\bibitem{uzan} J.P. Uzan, Rev. Mod. Phys.\textbf{\ 75}, 403 (2003): J.-P.
Uzan, astro-ph/0409424.

\bibitem{olive} K.A. Olive and Y-Z. Qian, Physics Today, pp. 40-5 (Oct.
2004).

\bibitem{BD} C. Brans and R.H. Dicke, Phys. Rev. \textbf{124}, 925 (1961).

\bibitem{Mota:2004} D. Mota, and J.D. Barrow, Phys. Lett. B, \textbf{581},
141 (2004); J.D. Barrow and D.F. Mota, Class. Quantum Gravity \textbf{20},
2045 (2003).

\bibitem{Mota:2003} D. Mota and J.D. Barrow, Mon. Not. Roy. Astron. Soc. 
\textbf{349}, 281 (2004).

\bibitem{marc} W.J. Marciano, Phys. Rev. Lett. \textbf{52}, 489 (1984).

\bibitem{Sandvik:2001} H. Sandvik, J.D. Barrow and J. Magueijo, Phys. Rev.
Lett. \textbf{88, }031302 (2002).

\bibitem{Bekenstein:1982} J.D. Bekenstein, Phys. Rev. D \textbf{25,} 1527
(1982).

\bibitem{two} J.D. Barrow, J. Magueijo and H. Sandvik, Phys. Lett. B \textbf{%
541,} 201 (2002).

\bibitem{Barrow:2002a} J.D. Barrow, H. Sandvik, and J. Magueijo, Phys. Rev.
D \textbf{65}, 063504 (2002).

\bibitem{Barrow:2002b} J.D. Barrow, H. Sandvik, and J. Magueijo, Phys. Rev.
D \textbf{65}, 123501 (2002).

\bibitem{Barrow:2002c} J.D. Barrow, J. Magueijo and H. Sandvik, Phys. Rev. D 
\textbf{66}, 043515 (2002).

\bibitem{Magueijo:2002} J. Magueijo J.D. Barrow and H. Sandvik, Phys. Lett.
B \textbf{549} 284 (2002).

\bibitem{Barrow:2001} J.D. Barrow and C. O'Toole, Mon. Not. R. astron. Soc. 
\textbf{322}, 585 (2001).

\bibitem{Kimberly:2003} D. Kimberly and J. Magueijo, Phys. Lett. B \textbf{%
584,} 8 (2004).

\bibitem{cham} N. Chamoun, S.J. Landau and H. Vucetich, Phys. Lett. B 
\textbf{504}, 1 (2001).

\bibitem{bass} D. Parkinson, B. Bassett and J.D. Barrow, 2003. Phys. Lett.
B, \textbf{578}, 235(2003).

\bibitem{mart} P.P. Avelino, C.J.A.P. Martins and J.C.R.E. Oliveira, Phys.
Rev. D \textbf{70}, 083506 (2004).

\bibitem{beknew} J.D. Bekenstein, Phys.Rev. D \textbf{66,} 024005 (2002).

\bibitem{dm} K.A. Olive and M. Pospelov, Phys. Rev. D \textbf{65,} 085044
(2002); E. J. Copeland, N. J. Nunes and M. Pospelov, Phys.Rev. D \textbf{69,}
023501 (2004); S. Lee, K.A. Olive and M. Pospelov, Phys.Rev. D \textbf{70,}
083503 (2004).

\bibitem{barrow:2004} Barrow, J.D., Kimberly, D. and Magueijo, J., Class.
Quantum Grav. \textbf{21}, 4289 (2004).
\end{thebibliography}
\end{document}